# ResBench: A Comprehensive Framework for Evaluating Database Resilience

Puyun Hu[†], Wei Pan[†*], Xun Jian[†], Zeqi Ma[†], Tianjie Li[†], Yang Shen[†], Chengzhi Han[†], Yudong Zhao[†], Zhanhuai Li[†]

[†]School of Computer Science and Engineering, Northwestern Polytechnical University
[†]Key Laboratory of Big Data Storage and Management, Northwestern Polytechnical University
[†]Dongxiang Road, Xi'an Shaanxi, China
hupy@mail.nwpu.edu.cn, {panwei1002, jianxun}@nwpu.edu.cn, {mazeqi, litj585, shenyan, hanchengzhi, zyd}@mail.nwpu.edu.cn, lizhh@nwpu.edu.cn

*Abstract*—Existing database benchmarks primarily focus on performance under ideal running environments. However, in real-world scenarios, databases probably face numerous adverse events. Quantifying the ability to cope with these events from a comprehensive perspective remains an open problem. We provide the definition of database resilience to describe its performance when facing adversity and propose ResBench, a benchmark for evaluating database resilience. This framework achieves automation, standardization, and visualization of the testing process through clear hierarchical decoupling. ResBench simulates adverse events and injects them during normal transaction processing, utilizing a module to gather multiple metrics for the evaluation model. We assess database resilience across eight dimensions: throughput, latency, stability, resistance, recovery, disturbance period, adaptation capability and metric deviation. All the results are presented to users via a user-friendly graphical interface. We demonstrate the execution process and result interpretation of ResBench using two types of adversity datasets.

## I. INTRODUCTION

Database benchmarks are typically used to quantify and compare the performance of databases. They generally consist of several key components such as datasets, workloads, testing programs and metrics. Benchmarks such as TPC-C, TPC-H, and TPC-DS primarily focus on transactional workloads, using pre-defined data schemas and operations to simulate real-world scenarios and measure system throughput. Driven by the emergence of new workload demands including HTAP and multi-model data processing, newer benchmarks like M2Bench [1] and HyBench [2] have been developed to evaluate databases from perspectives such as hybrid workload performance and multi-model support. However, they don't account for the comprehensive performance in complex scenarios involving adverse events. As real-world application environments are complex, they become inadequate in characterizing the database's ability to cope with adverse conditions and events.

The concept of system resilience [3] is often used to describe the degree to which a system can protect itself and its continuity-related assets from adverse events and conditions. Researchers believe that ensuring system resilience typically involves processes such as detection, resistance, recovery, and learning. In recent years, some benchmarks have attempted to evaluate some of these processes. For example, DBPA [4] has developed a transactional database anomaly benchmark to improve the accuracy of anomaly detection ML models and algorithms. CloudyBench [5] has designed a scoring system to evaluate cloud-native database capabilities. For the fail-over dimension, it monitors the time from the start of resistance to the lowest point and from the start of recovery to the end of recovery, calculating the impact and recovery capability scores. Meanwhile, there also exist many terms to assess the ability to cope with adversity, including elasticity, availability and robustness. However, these definitions all have limitations. For example, availability and elasticity focus on fault tolerance, typically indicating the ability to continue functioning despite failures or load fluctuations. However, failures and load fluctuations are only a subset of adverse events, which also include environmental changes and security threats. Robustness describes the ability to withstand disturbances, but does not address the recovery process. Despite these efforts, there is still no precise definition of resilience in the database field nor a complete benchmark to assess performance when facing adversity.

In this demo, we introduce the concept of system resilience to the database field for the first time. We give a comprehensive definition and eight evaluation dimensions of database resilience. These dimensions encompass traditional metrics such as throughput, latency, and stability, as well as other metrics including resistance, recovery, and adaptability. Based on this definition, we develop a flexible and extensible benchmark, ResBench, for comprehensively evaluating database resilience. In a nutshell, ResBench has the following key features:

**Abstract resilience patterns and real adversity datasets covering various adverse events.** ResBench can automatically locate the resilience triangle [6] reflecting adversity disturbance from the original performance and abstract five standard resilience patterns according to the shapes. At the same time, we use chaos engineering tools and load generators to construct a variety of fault and abnormal scenarios as the real adversity data, then combine them with abstract resilience patterns as the inputs of the benchmark. Therefore, ResBench supports diversified test schemes from basic faults to mixed adverse scenarios.

**A Pluggable, generalizable and customizable resilience testing framework.** ResBench includes execution layer, adapter layer, core layer, computing layer and GUI. The execution layer integrates load generation tools, adverse events injection tools, and monitoring tools. The highly pluggable adapter layer seamlessly integrates various tools through standard interfaces, allowing users to select the most needed tool. At the same time, ResBench has strong generalization capabilities. Its core programmable adversity simulation is independent of the database under test, which can easily adapt to different types of databases.

**Comprehensive resilience evaluation.** We combine the concept of the resilience triangle(which is always used to evaluate the resilience of engineering systems) with database characteristics. As a result, we build an eight-dimensional evaluation system. It comprehensively defines the database resilience through the entire process of pre-detection, in-process resistance and recovery, and post-event learning. Finally, results are presented to users in the form of radar charts, allowing different users to evaluate the resilience from different dimensions.

## II. DEFINITION OF DATABASE RESILIENCE

### A. Resilience Triangle Model

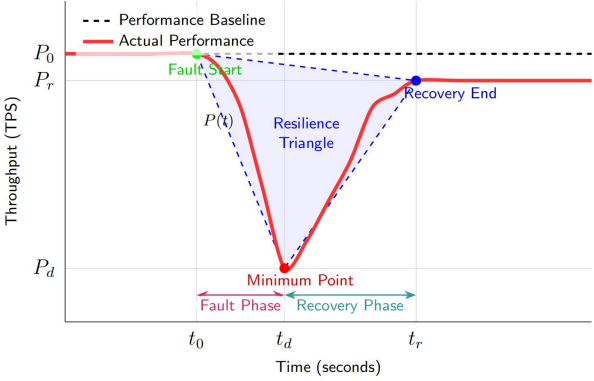

Fig. 1. Database resilience triangle

The resilience triangle model is proposed by Bruneau and others [6]. It represents the geometric region formed by the change curve and the time axis. Similarly, we first construct a change curve with time as the horizontal axis and throughput as the vertical axis. We use filters to smooth and reduce noise on the original curve, and define the final curve as $P(t)$. The database resilience triangle is constructed from the starting point of degradation located at $(t_0, P_0)$, the lowest point located at $(t_d, P_d)$ and the point of recovery completion located at $(t_r, P_r)$, as shown in Figure 1.

### B. Database Resilience Definition

We introduce the resilience concept to the database field and define it as follows: Database resilience is the ability that the database itself can ensure its key functions continue to provide service within a defined operating state when facing many adverse events, and can adaptively recover after these events disappear. We set hyperparameters including $\alpha$, $\beta$, $\gamma$ according to the specific scenarios. In ResBench, database resilience consists of the following dimensions:

(1) **Throughput Score**. This score represents the percentage of throughput loss through the process. We map percentages to values not greater than 100, which is similar to other scores.

$$S_{Tho} = 100 \times \frac{\int_{t_0}^{t_r} P(t)dt}{P_0 \times (t_r - t_0)} \quad (1)$$

(2) **Latency Score**. We define the latency curve as $L(t)$ and the baseline latency curve as $TL(t)$. This score reflects the percentage of latency amplification.

$$S_{Lat} = 100 \times \frac{\int_{t_0}^{t_r} TL(t)dt}{\int_{t_0}^{t_r} L(t)dt} \quad (2)$$

(3) **Stability Score**. This score reflects database stability. We normalize the data and divide them into $K$ segments. We calculate the variance of each segment, average these variances, and finally perform numerical mapping.

$$S_{Sta} = 100 \times \left(1 - \left(\frac{\frac{1}{K} \times \sum_{j=1}^{K} Var(P_{norm}(t))_{t \in I_j}}{0.25}\right)^{\alpha}\right) \quad (3)$$

(4) **Resistance Score**. This score reflects resistance. The first term describes resistance time, the second term describes absolute performance loss, and the third term describes the transaction volume loss.

$$S_{Res} = 100 \times \left(\omega_1 \times e^{-\beta \times (t_d - t_0)} + \omega_2 \times \frac{P_d}{P_0} + \omega_3 \times \frac{\int_{t_0}^{t_d} P(t)dt}{P_0 \times (t_d - t_0)}\right) \quad (4)$$

(5) **Recovery Score**. Similar to the resistance, this score describes recovery time, absolute performance loss, and the amount of transaction loss during the recovery phase.

$$S_{Rec} = 100 \times \left(\omega_1 \times e^{-\beta \times (t_r - t_d)} + \omega_2 \times \frac{P_r}{P_0} + \omega_3 \times \frac{\int_{t_d}^{t_r} P(t)dt}{P_0 \times (t_r - t_d)}\right) \quad (5)$$

Then perform a single test n times and calculate the following three multiple test scores in sequence:

(6) **Period Score**. This score represents the average period. We calculate the average duration of these tests.

$$S_{Per} = 100 \times \frac{\sum_{i=1}^{n} e^{-\beta \times (t_r^i - t_0^i)}}{n} \quad (6)$$

(7) **Deviation Score**. This score reflects the deviation. We define $S \in \{S_{Tho}, S_{Lat}, S_{sta}, S_{Res}, S_{Rec}\}$. For each type of the five scores, we sum the difference between the i-th result and the first result, then calculate the weighted arithmetic mean to measure the scores' fluctuation.

$$S_{Dev} = 100 \times \frac{\sum_{j=1}^{5}\left(\omega_j \times \sum_{i=1}^{n}\left(S^i - S^1\right)\right)}{5 \times n} \quad (7)$$

(8) **Adaptability Score**. This score reflects adaptability. We calculate the dispersion of scores relative to the mean.

$$S_{Ada} = 100 \times \sum_{j=1}^{5}\left(\omega_j \times \frac{\sum_{i=1}^{n}|S^i - \overline{S}|}{n \times \overline{S}}\right) \quad (8)$$

## III. RESBENCH DESIGN AND IMPLEMENTATION

### A. ResBench Architecture

In this section, we present the architecture of ResBench, as shown in Figure 2. It provides execution layer, adapter layer, core layer, computing layer and GUI. The functions of each layer are as follows:

**Execution Layer.** This bottom layer provides basic support. It integrates three components including WorkloadService, ChaosService and MonitorService. First, WorkloadService applies a continuous and stable load to the database under test using BenchmarkSQL [7] or other tools. ChaosService then injects various pre-set adverse events into the database. Some of them are simulated using ChaosBlade [8], while others run Linux commands. Finally, during the test, MonitorService continuously captures key metrics from the database. The collected metrics include system resource metrics and database external performance metrics, which serve as the core basis for evaluating database resilience.

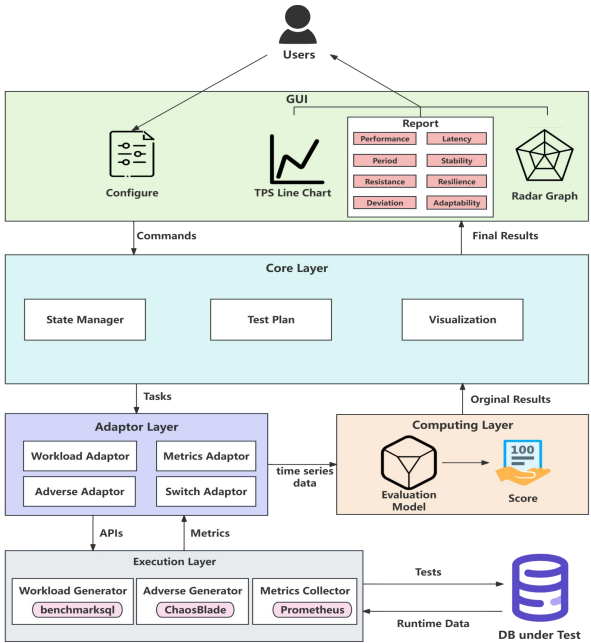

Fig. 2. The architecture of ResBench

**Adapter Layer.** The adapter service layer is the driver of ResBench. It encapsulates various standardized interfaces, enabling the core layer to centrally dispatch external tools. This layer decouples ResBench from auxiliary tools and the database under test, facilitating scalability. It includes SwitchAdapter, WorkloadAdapter, AdversityAdapter, and MetricsCollector. The SwitchAdapter is responsible for connecting and closing the database before and after testing, as well as managing its status. It encapsulates connection protocols for various databases. The WorkloadAdapter then converts abstract test instructions issued by the core layer into specific call commands or APIs, passing them to the execution layer. The AdversityAdapter translates abstract adverse scenarios defined by the core layer into specific injection commands. Finally, the MetricsCollector retrieves the required metrics from the execution layer based on test requirements, converts them into a unified data model, and passes them to the business layer for subsequent analysis.

**Core Layer.** This layer is the center of ResBench. It is responsible for receiving instructions from the GUI, coordinating the execution sequence of underlying components accordingly, and managing the test lifecycle. It serves as a logical hub that includes StateManager, TestScheduler, and Visualization converter. StateManager maintains global configuration information and monitors database connection status. TestScheduler organizes users' abstract test plans into several steps and controls the corresponding adapters. It tracks the execution status of each adapter throughout the whole process, catches exceptions, and provides corresponding feedback. After receiving data passed by the computing layer, the Visualization converter changes them into user-friendly visualizations.

**Computing Layer.** We have specialized the concept of resilience triangle to the database. In concrete terms, after collecting data from tests, the computing layer first preprocesses them, including smoothing, noise reduction, and filling in missing points. Then, using detection algorithms, we identify consensus-based change-points. Three key points are identified from these points: the starting point of degradation, the lowest point, and the point of recovery completion. Based on these points, we construct a resilience triangle describing the adversity response process. The relevant data are then fed into the evaluation model, which calculates scores and uploads results to the core layer.

**GUI.** We design a comprehensive and user-friendly GUI for users to interact with ResBench. The diagrams are shown in Figure 3. It provides users with friendly graphical forms, allowing users to visually configure complex adverse test scenarios without writing code. It includes functions such as database connection, workload configuration and adversity selection, as shown in ①,②,③. In addition, it also provides a query function for historical records, as shown in ④.

### B. Description of core tools

Our framework relies on the collaborative work of three major external components: load generators, adverse events injection tools and monitoring tools. ResBench uses the following three mature tools:

**Workload Generation Tools.** Workload generation is the foundation of the framework. It issues queries to the database mimicking the real production workload. BenchmarkSQL [7] is an open-source relational benchmark tool. It simulates an order transaction scenario and supports various configurations. We use it to generate the sustained and stable TPC-C workload.

**Adverse Events Injection Tools.** ResBench currently supports adverse events on CPU, memory, network and I/O. To systematically inject them, ResBench utilizes Chaosblade [8], Alibaba's open-source chaos engineering tool. It has some basic fault injection functions, and we make targeted improvements to it. We use this tool to construct typical failures such as CPU saturation, I/O congestion, network packet loss and others. However, it must be noted that ChaosBlade [8] primarily focuses on failures, which are only one component of adverse events. For this reason, this module is not tightly bound to ChaosBlade [8] or any single tool. Instead, we design it as a pluggable interface. This means that custom adapters can be written based on testing requirements to integrate other tools or directly use Linux commands. This design ensures comprehensive coverage of various adverse scenarios and provides powerful flexibility.

**Metrics Collecting Tools.** To quantitatively assess database state changes, continuous collection of system metrics is also essential. Prometheus [9] is an open-source system monitoring software. It periodically captures metrics and stores them in an efficient time series database. During resilience testing, we use it to continuously captures changes in database throughput and transaction latency, which providing support for the calculation of resilience score.

### IV. DEMONSTRATION

We use two datasets to demonstrate ResBench's testing process and results. One is a real adversity dataset including several common adverse events, and the other is an abstract dataset consisting of five standard resilience triangles.

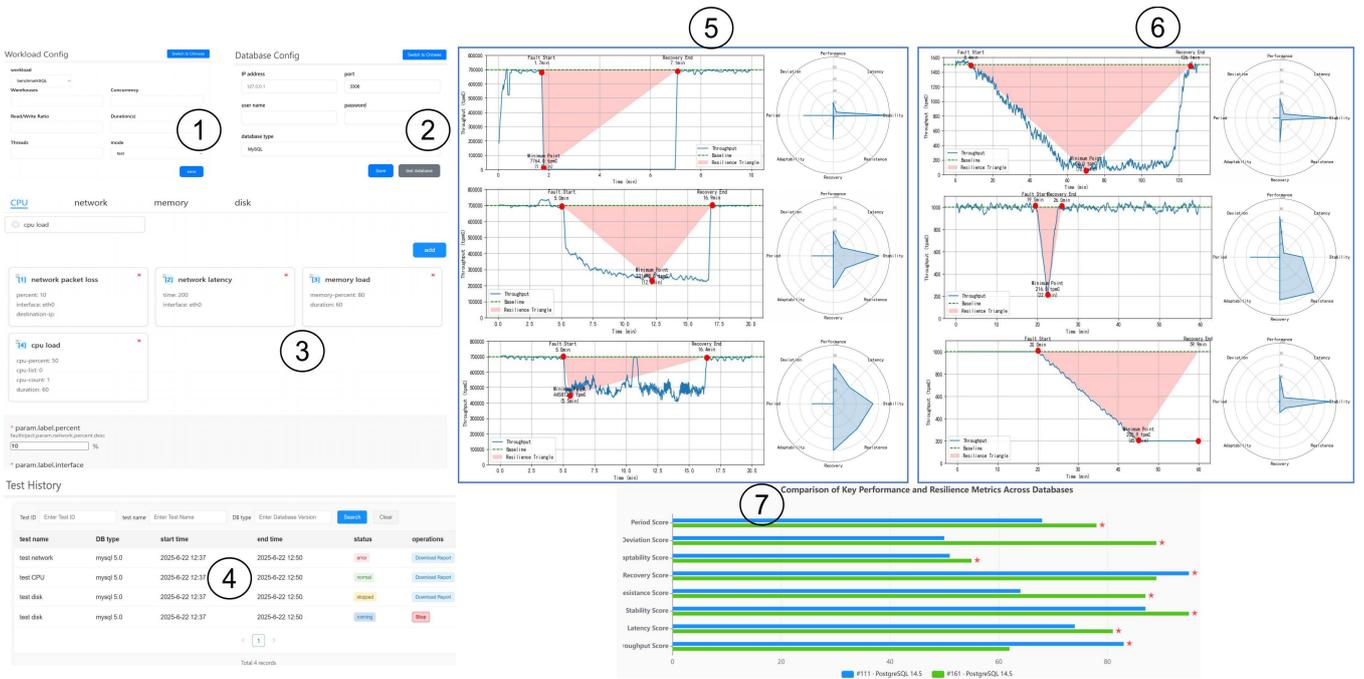

Fig. 3. ResBench GUI and results

*A. Real Adversity Dataset*

We first show some real adversity test results. We apply a stable TPC-C workload to a commercial database. After its performance stabilized, we use ChaosBlade [8] to inject three common adverse events, including network packet loss, high CPU load and high I/O load. The results are shown in ⑤ from the top down. The radar charts show that the database exhibits a similar U-shaped trend for all three types of adverse events, and the scores for each event are inversely proportional to the degree of decline. These conclusions can provide an effective basis for users to select the required database when they receive similar feedback.

*B. Abstract Dataset*

Although the above adverse events all present the U-shaped triangle, in order to test the theoretical expression ability of the ResBench, we abstract five types of resilience patterns with different shapes:

(1)**V-shaped triangle**, representing the normal recovery after performance degradation.

(2)**U-shaped triangle**, representing the period of being trapped in low throughput and then normal recovery

(3)**Partial recovery triangle**, representing permanent damage to the database under adverse impact, which results in the inability to fully recover to the original throughput.

(4)**Recovery failure triangle**, representing the severe impact on the database, the failure of the recovery mechanism, and the complete loss of service capabilities.

(5)**Recovery overshoot triangle**, representing database triggering optimization, breaking through the bottleneck and improving its throughput to a higher level.

Next, we generate five copies of each type and rank them from worst to best, then directly upload them to the computing layer. ⑥ in Figure 3 shows ResBench's resilience triangles and radar charts for V-shaped, U-shaped and recovery failure triangle one by one. ResBench considers the U-shaped triangle to be relatively stable, the V-shaped triangle to have a relatively strong resistance, and the recovery failure triangle to have an extremely low recovery. For multiple results, users can compare them in the GUI. For example,⑦ shows the comparison result of the first and last test of the V-shaped triangle. In fact, period, deviation and adaptation scores become higher. These results are generally consistent with subjective judgment. This demonstrates that ResBench can accurately analyze resilience patterns and provide specific results for intuitive reference.